# Fine-Tuning Bulk-oriented Universal Interatomic Potentials for Surfaces: Accuracy, Efficiency, and Forgetting Control

Jaekyun Hwang1, Taehun Lee2, Yonghyuk Lee3,*, Su-Hyun Yoo1,*

1. Digital Chemical Research Center, Korea Research Institute of Chemical Technology, Daejeon 34114, Republic of Korea

2. Division of Advanced Materials Engineering, Jeonbuk National University, Jeonju 54896, Republic of Korea

3. Department of Chemical and Biochemical Engineering, Dongguk University, Seoul, 04620, Republic of Korea

* yhyuk@dongguk.edu, syoo@krict.re.kr

## Abstract

Accurate prediction of surface energies and stabilities is essential for materials design, yet first-principles calculations remain computationally expensive and most existing interatomic potentials are trained only on bulk systems. Here, we demonstrate that fine-tuning foundation machine learning potentials (MLPs) significantly improves both computational efficiency and predictive accuracy for surface modeling. While existing universal interatomic potentials (UIPs) have been solely trained and validated on bulk datasets, we extend their applicability to complex and scientifically significant unary, binary, and ternary surface systems. We systematically compare models trained from scratch, zero-shot inference, conventional fine-tuning, and multi-head fine-tuning approach that enhances transferability and mitigates catastrophic forgetting. Fine-tuning consistently reduces prediction errors with orders-of-magnitude fewer training configurations, and multi-head fine-tuning delivers robust and generalizable predictions even for materials beyond the initial training domain. These findings offer practical guidance for leveraging pre-trained MLPs to accelerate surface modeling and highlight a scalable path toward data-efficient, next-generation atomic-scale simulations in computational materials science.

## Introduction

Machine learning potentials (MLPs) are data-driven computational models designed to accurately predict potential energy surfaces from given atomic configurations. The primary advantage of MLPs is their ability to integrate the computational efficiency of classical empirical interatomic potentials and force fields with the predictive accuracy of first-principles methods, notably density functional theory (DFT) [1-3]. While conventional DFT calculations typically scale as $O(N^3)$ for solid-state systems, MLPs drastically reduce computational costs to $O(N)$, thereby enabling simulations that were previously impractical, such as high-throughput screening and large-scale molecular dynamics on nanometer and nanosecond scales [4].

Nevertheless, the practical effectiveness of MLPs is often limited by the scarcity of publicly available comprehensive training datasets. This limitation frequently compels researchers to generate custom datasets tailored for individual studies. Although repositories such as Interatomic Potentials Repository offer pre-trained potentials, their applicability is constrained by the elemental compositions and structural configurations originally included, which can result in significant performance degradation when applied beyond their original training domains [5-7].

Recent advances have led to the development of Universal Interatomic Potentials (UIP), notably enabled by E(3)-equivariant graph neural network architectures such as NequIP and MACE [8,9]. UIPs leverage rich geometric representations to achieve high accuracy and improved data efficiency across diverse material classes, enabling the simultaneous prediction of interactions among nearly all elements [10]. Current UIP models are predominantly trained on bulk material datasets and primarily benchmarked on their ability to predict the stability and energetics of crystal structures [11].

While accurate bulk crystal structure predictions are important, many practical applications—such as chemical reaction modeling, and thin-film deposition—demand reliable simulations of surfaces and interfaces. The surface energy of materials is a fundamental physical property that influences a wide range of physicochemical processes such as heterogeneous catalysis, corrosion, and related ones. However, systematically curated databases explicitly tailored for surface systems remain limited, constraining the development of reliable surface-specific MLPs. Furthermore, benchmark studies have shown that UIPs trained exclusively on bulk datasets tend to systematically underestimate surface energies, a phenomenon known as softening [12].

One effective approach to overcome such issue is fine-tuning, a widely adopted strategy in deep learning. Recent tutorial-style work has outlined fine-tuning workflows for UIP, covering case studies on solid electrolytes, stacking faults, and interfacial systems [13]. Fine-tuning refines a pre-trained foundation model with targeted additional data to enhance predictive accuracy in a specific domain. This approach has achieved notable success across multiple fields, including image recognition and natural language processing [14,15]. When applied to UIPs, fine-tuning can reduce the amount of training data required to reach a given accuracy [16], and an emerging work show that a model fine-tuned on the MD trajectory of one Li solid-state electrolyte can be reused for MD simulations of other Li electrolytes [17]. Despite these benefits, fine-tuning can cause catastrophic forgetting, in which newly acquired knowledge disrupts previously learned representations [18]. In domains with abundant data, traditional fine-tuning approaches, like adjusting learning rates and training epochs, typically suffice. In contrast, data-scarce or specialized settings often require more advanced methods, such as multi-head fine-tuning or continual learning frameworks, to preserve general knowledge while incorporating new domain-specific information [17,19,20]. Despite these advances, to the best of our knowledge there has been no systematic study that applies and rigorously evaluates fine-tuning of UIPs on out-of-training-domain systems—specifically surfaces—under controlled comparisons with baselines. We address this gap by conducting a comprehensive, dataset-controlled assessment of fine-tuning and multi-head fine-tuning for surface systems.

In the context of UIPs, multi-head fine-tuning entails training the model concurrently on the original bulk datasets and newly introduced surface datasets. This technique helps the model maintain its bulk prediction capabilities while enhancing surface-specific accuracy. Inspired by multi-task learning, Multi-head fine-tuning reduces catastrophic forgetting by allowing models to maintain pre-trained knowledge, promoting positive transfer during continual learning [21, 22]. Despite these advantages, adopting a multi-head architecture inherently increases both model complexity and computational cost.

This study aims to systematically evaluate the effectiveness of MLPs in predicting surface properties of materials through three strategies: (1) training a MLP model from scratch on surface data (denoted in the following as *TS*), (2) employing pre-trained UIPs without modification, i.e., zero-shot models (*ZS*), and (3) fine-tuning UIP foundation models (*FT*). We specifically investigate improvements in predicting surface energetics relevant to surface reactions, 2D materials, and nanoscale phenomena with three distinct systems (unary, binary, ternary). In addition, we explore the effect of multi-head fine-tuning model (*MHFT*) to determine whether simultaneous training on both bulk and surface datasets mitigates catastrophic forgetting and enhances predictive accuracy. The overall workflow is schematized in Figure 1, which contrasts bulk and surface datasets and highlights the distinct initialization strategies across models. To our knowledge, this study provides an initial systematic assessment of fine-tuning UIPs for material surfaces that lie outside the original bulk-trained domain. Ultimately, this study is expected to provide clear guidelines for selecting a strategy that extend UIPs beyond bulk materials.

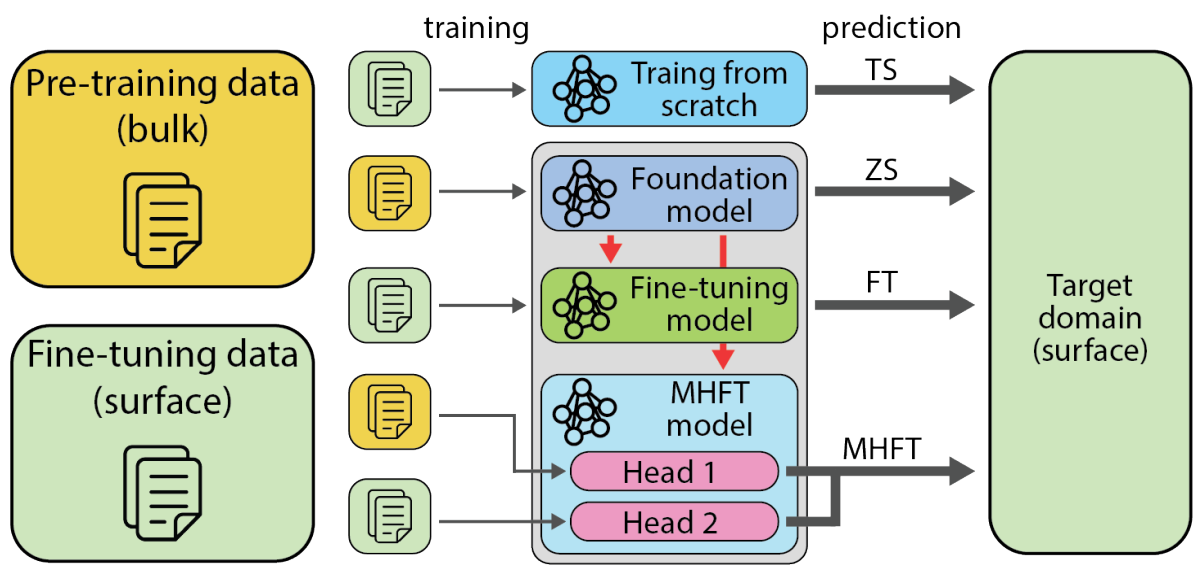

**Figure 1.** Graphical abstract of the training schemes for the predictive models used in this study. Yellow rectangles denote large-scale bulk datasets, and green rectangles surface datasets, which constitute the prediction target. The foundation model and the *TS* model are trained from random initialization, whereas the *FT* and *MHFT* models are initialized from the foundation model.

**Results**

Before assessing the aforementioned strategies, we benchmarked several pre-trained UIP models, such as M3GNet[10], SevenNet-l3i5[23], MACE-MP-0a-small[9,24], ORBv2[25], and eqV2-M[26], on diatomic potential energy curves to evaluate their robustness and generalizability. Since diatomic scans directly probe the short- and long-range behavior of the learned interactions, they provide an intuitive and physically transparent sanity check of model transferability. While most models reproduced equilibrium bond lengths, some exhibited irregularities, whereas MACE-MP-0a-small (hereafter, "MACE") provided physically reliable behavior. Considering performance and implementation practicality, we selected MACE as the foundation model for subsequent calculations (detailed in Supplementary Notes 1).

**Case 1. Unary System: Gold (Au) Surface**

We first investigated a simple unary system using computational data for gold (Au) surfaces. Gold is widely utilized in catalysis due to its unique surface chemical properties [27]. Notably, Au is an FCC metal whose (111) surface reconstructs at room temperature [28]. However, modeling such long-range surface reconstruction typically demands simulation cells that exceed practical computational limits of standard DFT calculations.

To address this computational challenge, Cameron et al. curated a comprehensive first-principles dataset of 2,545 gold surface configurations to construct their MLP [29]. They systematically and sequentially selected calculation targets to maximize the diversity of surface configurations. Because configurations generated later in the workflow are likely challenging to predict from earlier data, we allocated the first 2,000 configurations (approximately 79%) for training and validation, and reserved the remaining 545 configurations (approximately 21%) as the test set.

Rather than optimizing data selection, we isolate the effect of *FT* by holding the training set fixed and contrasting it with *TS*. A systematic study on how different data selection strategies affect model performance and efficiency is left for future work. Accordingly, we followed the sequential selection protocol used by the original datasets.

Figure 2(a) compares the prediction errors of *ZS*, *TS*, and *FT* models as the training data size increases (i.e., 40, 200, and 2,000 configurations for both *TS* and *FT*). A zero-shot prediction scenario is presented as a baseline for fine-tuning results. In all cases, larger training datasets reduced the prediction error, and *FT* models consistently outperformed *TS* models, indicating the benefit of incorporating prior knowledge from the foundation model.

The *ZS* scenario demonstrated reasonably good predictive performance for gold surfaces even without fine-tuning, exhibiting energy prediction errors of 14.8 meV/atom and force prediction errors of 74.7 meV/Å. Interestingly, even minimal fine-tuning with 40 data points significantly reduced prediction errors to 4.3 meV/atom and 37.4 meV/Å for energy and forces, respectively. Remarkably, the *TS* model on 2,000 configurations achieved similar or worse performance compared to *FT* models trained with just 40–200 configurations. This demonstrates the efficiency and effectiveness of fine-tuned models for targeted systems.

The advantage of the *FT* model is further evident when examining the Au–Au diatomic potential shown in Figure 2(b). Here, we compared potentials calculated directly from first-principles calculations to predictions made by *TS* and *FT* models, both models were trained on a 2,000 configurations. The scenario involving two isolated atoms in vacuum fundamentally differs from investigating energetically stable surface structures. Nevertheless, this scenario serves as a valuable measure of how effectively the models extrapolate to unexplored out-of-domain (OOD) configurations.

Both models exhibited similar behaviors near the equilibrium distance (~2.5 Å). However, the *TS* model exhibited physically unrealistic behaviors, such as unstable and irregular curves at short distances (< 1.8 Å), as well as incorrect long-range repulsive interactions. In contrast, the fine-tuned model provided significantly more accurate descriptions across the entire energy landscape. To investigate the origin of these discrepancies, we computed the Au–Au radial distribution function (RDF) of the training data in Supplementary Fig. 3. No Au–Au pairs closer than 2.21 Å were present, and the first RDF peak appears at 2.91 Å. The absence of energetically unstable diatomic configurations shorter than 2.2 Å in the training dataset is likely responsible for the poor predictions of short-range interactions. *TS* model receives no information to correct the steep energy well at near-collision distances. Consequently, it is possible that the model does not sufficiently account for highly unstable states in which the two atoms are in extremely close proximity. Even beyond 3.5 Å, the *TS* model also incorrectly predicted repulsive interactions in the long-range regime. These inaccuracies likely stem from the limited representation of long-range configurations inherent in slab-based training data, compounded by residual interactions between periodic images, which together hinder accurate extrapolation beyond the sampled distance range.

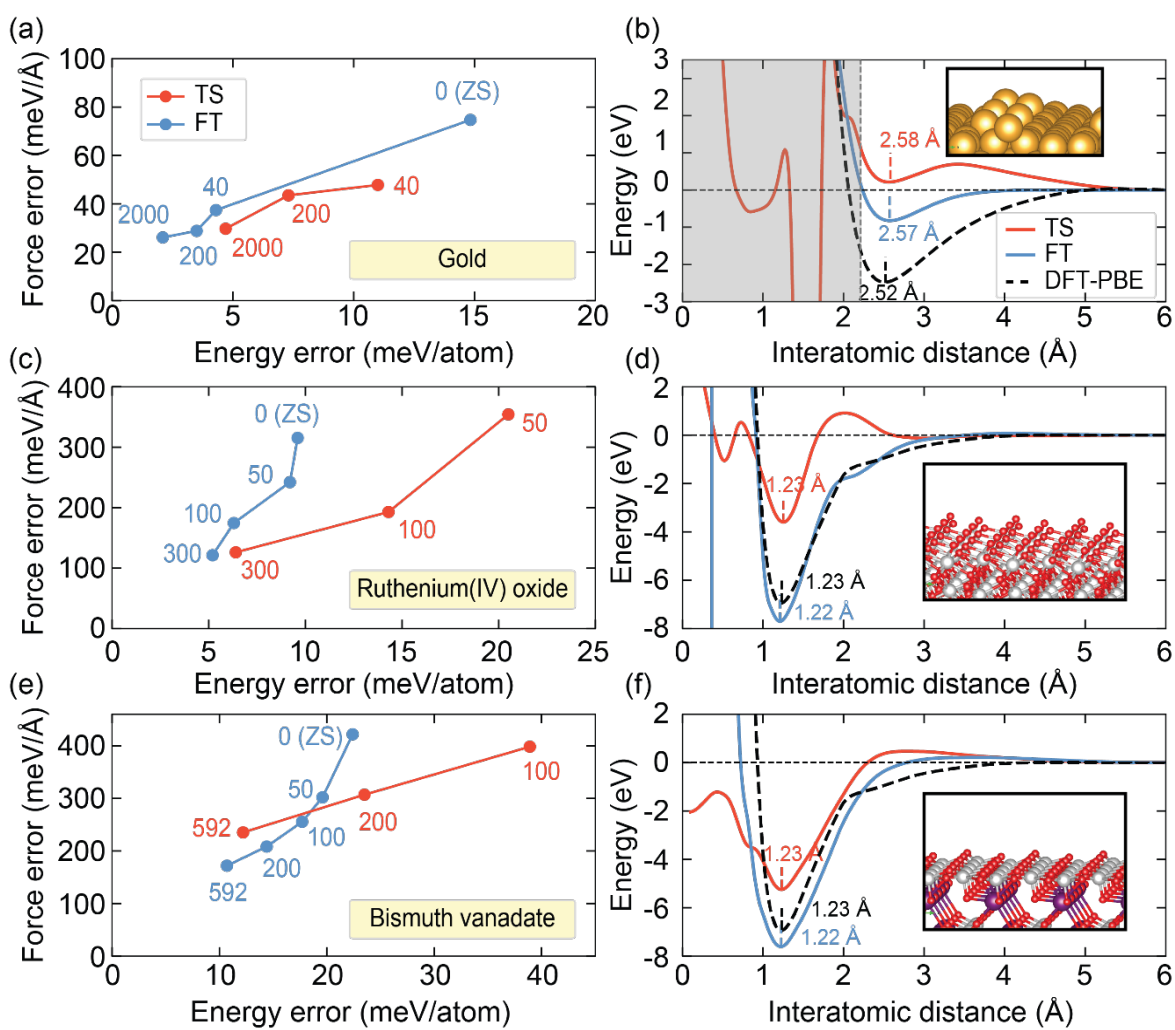

**Figure 2.** Training results for three surface systems in this work: (a-b) gold, (c-d) ruthenium(IV) oxide, and (e-f) bismuth vanadate. (a, c, e) Root mean squared errors (RMSE) for energies and forces evaluated on the test set as a function of training data size. Red (blue) lines correspond to *TS* (*FT*) models with random initialization. Annotated numbers adjacent to the points denote training dataset sizes, with "0" indicating *ZS* predictions. (b, d, f) Diatomic potential energy curves predicted by the *TS* and *FT* models trained on the largest dataset, compared with DFT-PBE reference calculations. The Au-Au pair potential for the gold surface is shown in (b) with a gray shading indicating the region where Au-Au bonding is absent. And O-O for the other two oxide surfaces in (d) and (f). In the right panel, short dashed vertical lines and labels indicate equilibrium bond lengths. Insets illustrate representative side-view atomic configurations of the studied surface systems.

**Case 2. Binary System: Ruthenium(IV) oxide (RuO$_2$) Surface**

Ruthenium(IV) oxide (RuO$_2$) surfaces have attracted significant interest due to their thermal stability and catalytic activity, particularly in heterogeneous catalysis and electrochemical energy applications such as the oxygen evolution reaction (OER). However, compared to simpler metallic surfaces, RuO$_2$ surfaces exhibit significantly greater complexity due to intricate surface reconstructions and dynamic interfacial phenomena [30–32]. In this study, we utilize first-principles data from Y. Lee et al., who extensively investigated various RuO$_2$ surface reconstructions using machine-learned potentials, to systematically train and validate our fine-tuning techniques [33].

The dataset for RuO$_2$ consisted of 340 calculated structures, slightly fewer than the gold surface case previously discussed. From this dataset, up to 300 structures (approximately 88%) were used as the training set and 40 structures (approximately 12%) as the test set. Figure 2(c) presents prediction errors as a function of the training dataset size. Similar to our observations for the gold surface, we found that increasing the size of the training dataset improved the predictive accuracy of all models. In all scenarios, assuming identical training data were used, the *FT* model consistently exhibited lower prediction errors compared to the *TS* model, highlighting the effectiveness of fine-tuning strategies.

However, a notable difference emerges when comparing the predictions for RuO$_2$ surfaces to those for gold surfaces, specifically regarding the magnitude of the force prediction errors, which are several times larger for RuO$_2$. This increased error likely reflects the more complex bonding and multiple atomic species in binary oxide systems, where factors like variable coordination environments, partial charge transfer, and stronger anisotropy make accurate force prediction more challenging. In addition, a recent study has shown that sharp features in the d-orbital density of states of transition metals can cause steep changes in the potential energy surface under small atomic rotations, further exacerbating prediction errors in models trained without sufficient coverage of such configurations [34].

Furthermore, similar to the gold case, we examined the validity of our models by analyzing pair potentials to represent interatomic interactions. For illustrative purposes, Figure 2(d) specifically shows the O–O pair potentials among three possible pairs (i.e., Ru-Ru, Ru-O, and O-O). A complete set of comparisons is provided in Supplementary Fig. 4. The *TS* model notably revealed three local minima occurring at distances of approximately 0.5 Å, 1.2 Å, and 3.0 Å. These features are clearly unphysical, indicating problematic behavior characterized by unexpected attractive interactions at distances below 0.7 Å. Such a profile poses a risk by potentially stabilizing non-physical configurations where oxygen atoms are unreasonably close.

In contrast, the *FT* model yielded a physically reasonable pair potential profile, exhibiting a primary local minimum at approximately 1.2 Å, consistent with the DFT reference. However, the *FT* model also showed an spurious attractive region at distances shorter than 0.4 Å. Despite this feature, the corresponding energy barrier is extremely high (~500 eV), which effectively prevents unphysical atomic configurations from arising during properly conducted MD simulations or structure exploration processes. Furthermore, incorporating short-range regularization terms or employing basis functions with enhanced resolution near the repulsive wall could mitigate such artifacts without significantly increasing computational cost [35, 36].

Moreover, previous studies employing MLPs in structural explorations have demonstrated that although possessing fully accurate pair potential is beneficial, such accuracy is not always strictly necessary for successful simulations [37]. Nonetheless, ensuring that the potentials remain physically reasonable significantly enhances model reliability and interpretability in computational predictions.

**Case 3. Ternary system: Results for Bismuth vanadate (BiVO$_4$) surface**

Bismuth vanadate (BiVO$_4$) is a widely studied photocatalyst for solar water splitting. Its surface plays a critical role, governing charge separation, carrier transport, and the oxygen evolution reaction (OER). A quantitative understanding of BiVO$_4$ surfaces is essential for optimizing photoelectrochemical performance and guiding the rational design of surface-engineered or heterostructured systems. In this context, BiVO$_4$ also serves as a challenging yet informative testbed for developing and testing UIPs in complex transition metal oxides. We used a dataset of 928 DFT surface calculations reported by Y. Lee et al. [38], dividing it into 592 configurations (approximately 64%) for training and validation, with the remaining 336 configurations (approximately 36%) reserved for testing. The test set was increased relative to the previous two cases to enable detailed analysis of outliers that did not appear previously, as discussed below.

Figure 2(e) demonstrates that *FT* models consistently outperform *TS* models, aligning with our previous findings in unary and binary systems. Our comparisons through the three cases clearly indicate that fine-tuning a pretrained model, even when initially trained on different material systems, results in more accurate predictions than traditional training methods without pretraining.

Additionally, as shown in Figure 2(f), the oxygen potentials predicted by the *TS* model exhibit significantly weaker short-range repulsion, suggesting a risk of atomic collisions during structural optimization. In contrast, the *FT* model predicts oxygen binding energies more accurately and provides a reasonable potential shape, clearly demonstrating superior performance. Supplementary Fig. 4 provides additional comparisons for all six possible pair potentials in the ternary $BiVO_4$ system alongside with those for $RuO_2$ system. Although quantitative comparison in accuracy between *TS* and *FT* models is challenging, the *TS* model consistently struggles across all pair potentials. The *TS* model demonstrates multiple local minima and abrupt, non-smooth energy curves at short distances. Conversely, while the *FT* model shows abnormal behavior at extremely short distances (below 0.6 Å) in Bi-Bi and Bi-O pairs, these regions still exhibit sufficiently large energy barriers, confirming its overall superiority.

While further investigating the cause of high prediction errors in the $BiVO_4$ system by examining atomic forces, we identified unique behaviors not observed in the unary or binary systems. Figure 3 compares calculated and predicted atomic forces in $BiVO_4$ test set across all the three prediction models, plotting absolute atomic forces on a log-log scale. The black dashed diagonal line represents perfect agreement. Most data points located within regions bounded by red and blue dashed lines represents prediction error of ±0.4 eV/Å and ±4 eV/Å, respectively. Regions with higher point density illustrated as green to yellow region.

Figure 3 provides several insights. Firstly, considering data points outside the blue dashed line, which represent prediction errors greater than 4 eV/Å, the *ZS* model surprisingly has ~~only~~ two outliers, indicating reasonable prediction performance. However, the *TS* model, despite its lower force RMSE (0.24 eV/Å compared to 0.42 eV/Å of the *ZS* model), has 13 outliers significantly more than the *ZS* model as shown in Figure 3(b). This gap indicates that the TS model well predicts many atoms with small forces but fails notably in predicting large forces (~10 eV/Å), likely due to insufficient training on unstable atomic environments.

In contrast, the *FT* model not only achieves the lowest force prediction error (0.17 eV/Å) but also reduces outliers to three, comparable to *ZS* but superior overall, making *FT* model the most reliable prediction model among the three. Although our foundation model was trained solely on bulk data, reference [9] Figure 59(b) shows it included many high-force (~100 eV/Å) data points. Thus, the *FT* model leverages the foundation model's knowledge of high-force environments to effectively suppress outliers.

While large proportional prediction errors in small force regions (e.g., 50% error at 0.1 eV/Å) might not severely impact simulations like surface optimization and molecular dynamics. However, such inaccuracies in high-force regions can significantly distort surface structure optimization results, e.g., by stabilizing unstable atoms along unphysical pathways. Thus, applying the *FT* model to highly uncertain simulation scenarios not encountered during training significantly reduces the risk of incorrect predictions.

Such complexities in force prediction were not observed in the two simpler surface systems (Au and $RuO_2$) studied previously. Also Figures 3(b) and (c) show symmetric distributions along the diagonal line, indirectly confirming unbiased predictions and validating the correctness of our training procedures. While additional training data on various atomic environments would naturally resolve these issues, *FT* approach is an effective safeguard for research constrained by limited computational resources.

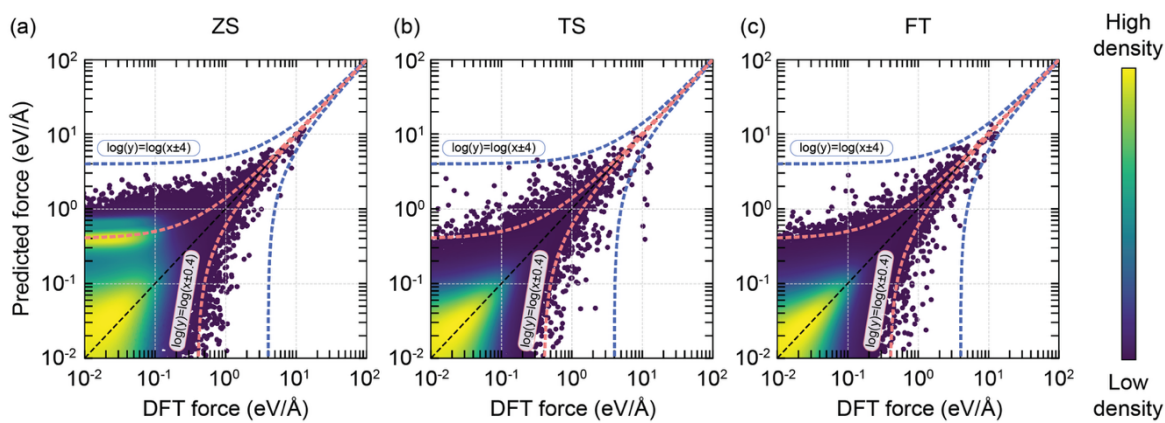

**Figure 3.** Comparison of DFT-calculated forces and predicted forces on atoms in the test dataset of the bismuth vanadate surfaces by three different models. Each point represents an absolute force vector on a single atom, with higher point densities shown in yellow. Dashed lines indicate deviation levels between predicted and DFT forces: the black dashed line represents perfect agreement, while the red and blue dashed lines correspond to deviations of ±0.4 eV/Å and ±4 eV/Å, respectively.

**Case 4. Impact of Multi-Head Fine-Tuning on Unary Novel Metal Surface Data**

In previous sections, we discussed the advantages and merits of *FT* models. However, *FT* also has its limitations. Because *FT* inherently modifies the parameters of the foundation model, predictions inevitably differ before and after *FT*. Moreover, since the fine-tuning data typically targets a specific surface system in this work, the fine-tuned model does not necessarily generalize well to other systems, particularly those well-represented in the original pre-training dataset. In extreme cases, the model may entirely lose its predictive performance for the pre-training data, a phenomenon known as "catastrophic forgetting" [18]. One method to mitigate catastrophic forgetting involves training simultaneously on both additional data and the original pre-training data [39,40]. *MHFT*, as implemented in MACE software package, combines loss functions calculated from both additional and original data, thereby maintaining accuracy across both datasets. Although this approach effectively prevents catastrophic forgetting, it significantly increases training time, as losses must be computed for the original pretraining dataset at every training epoch. *MHFT* requires roughly 100 times more training time than *FT*, as our pretraining dataset is 100 times larger than the FT dataset.

*MHFT* aims to retain pre-training competence and improve accuracy in the fine-tuned domain. To investigate the effectiveness of *MHFT* in scenarios similar to the gold surface data explored in Case 1, we utilized a computational database of various unary elemental surfaces published by Richard Tran et al. [41]. We hypothesized that a model appropriately fine-tuned on gold surface data would also exhibit reasonably accurate predictions for similar novel metal elemental surfaces. Contrary to our expectation, a model fine-tuned exclusively on gold surface data exhibited poorer predictive accuracy for other unary elemental surfaces than the zero-shot predictions of the original pre-trained foundation model. Ideally, if the *FT* model had effectively captured the generalizable characteristics of surface energies, it would also be expected to perform well on out-of-domain (OOD) unary metal surfaces. To quantitatively evaluate this problem, we specifically selected four elements—Ag, Cu, Pt, and Ir—due to their proximity to gold in the periodic table and their anticipated similar properties.

We applied *MHFT* using gold surface data (2,000 datapoints). To reduce training time, we randomly selected 200,000 data points from the complete MPTrj [42] dataset, which consists of approximately 1.5 million entries. Figure 4 compares the percentage prediction errors of surface energies for the four novel metals using three models: the ZS model (the original MACE foundation model), the *FT* model on gold surface data, and the *MHFT* model on gold surface data.

All three models predict total system energies consistent with the first-principles calculations, which were subsequently converted into surface energies based on the supercell geometry. Figure 4 illustrates the distribution of percentage prediction errors in surface energies for the selected metals (Ag, Cu, Pt, Ir), excluding gold, the element used for training. The RMSE of each model is 27, 88, and 16 meV/Å$^2$, respectively. Supplementary Fig. 5 shows same plot by absolute prediction errors.

Despite the limited dataset size, the *MHFT* model clearly demonstrated superior generalizability compared to the *FT* model and exhibited improved predictive accuracy relative to the zero-shot model, thereby indicating stronger performance in OOD scenarios. Analyzing the prediction error trends for each data point, we observed that the zero-shot model tended to slightly underestimate surface energies as expected [12], whereas the *FT* model consistently and significantly underestimated them in nearly all cases. The *MHFT* model exhibited an overall improvement, reducing prediction errors compared to the zero-shot model. Quantitatively, the RMSE of the *MHFT* predictions was nearly half that of the zero-shot predictions.

However, comparing the *MHFT* results with the results from Case 1 (summarized in Table 1), we observed a slight degradation in prediction accuracy for the specific gold surface data used in additional training. This decrease in accuracy might be explained by the fact that the gold surface prediction error remains an order of magnitude lower than the pre-training data prediction error. Although the energy error for gold increased by more than a factor of two (from approximately 1.3 meV/atom to 2.8 meV/atom), this difference remains negligible when compared to the typical reproducibility limit of DFT calculations (approximately 1–2 meV/atom) [43].

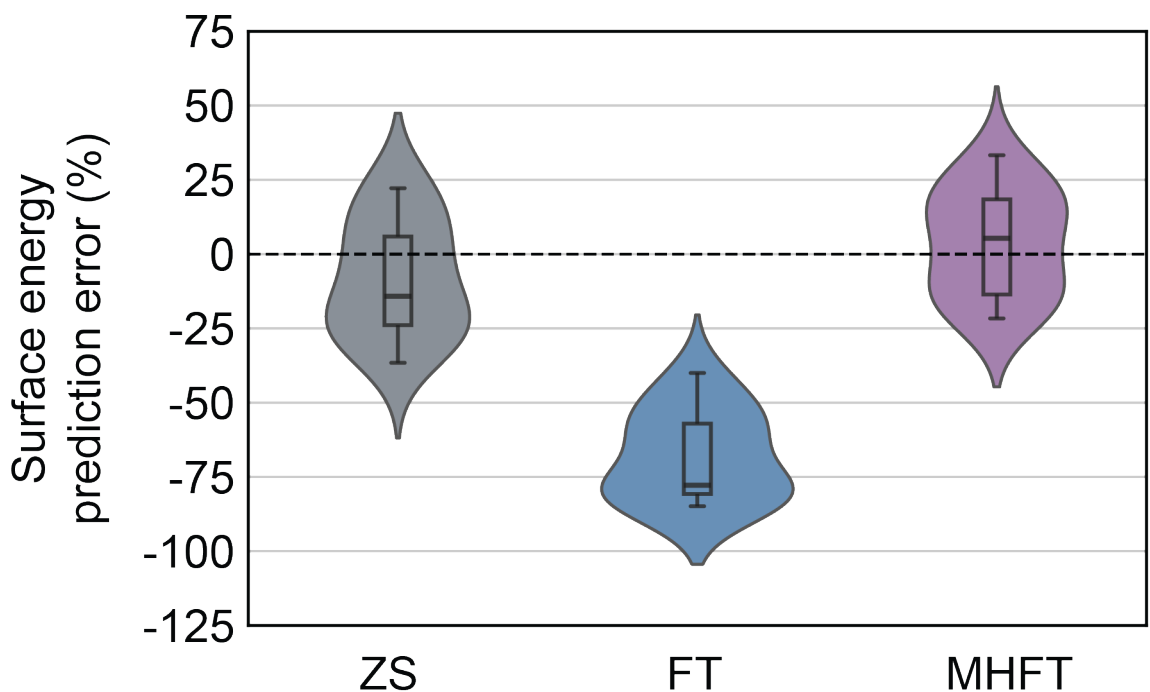

**Figure 4.** Violin plots of prediction errors in surface energy for surfaces of four noble metal elements (Ag, Cu, Pt, Ir) from three different models. Training was performed using 2,000 Au surfaces for *FT* and *MHFT* models, while all predictions were performed on 54 surfaces of Ag, Cu, Pt, Ir, excluding Au. The areas of the violins are normalized to be equal.

|  | MPTrj bulk data [42] | | Gold surface data [29] | |
| --- | --- | --- | --- | --- |
|  | Energy error | Force error | Energy error | Force error |
| Foundation model trained by MPTrj bulk data | **11 meV/atom** | **73 meV/Å** | 12 meV/atom | 55 meV/Å |
| + Fine-tuning | 302 meV/atom | 216 meV/Å | **1 meV/atom** | **21 meV/Å** |
| + Multi-head fine-tuning | 27 meV/atom | 111 meV/Å | 2.8 meV/atom | 28 meV/Å |

**Table 1.** Comparison of prediction errors on MPTrj bulk dataset (used to train the foundation model) and the gold surface dataset (used for fine-tuning), evaluated across three prediction models. Boldface represents the model and results that achieve the best performance on each dataset.

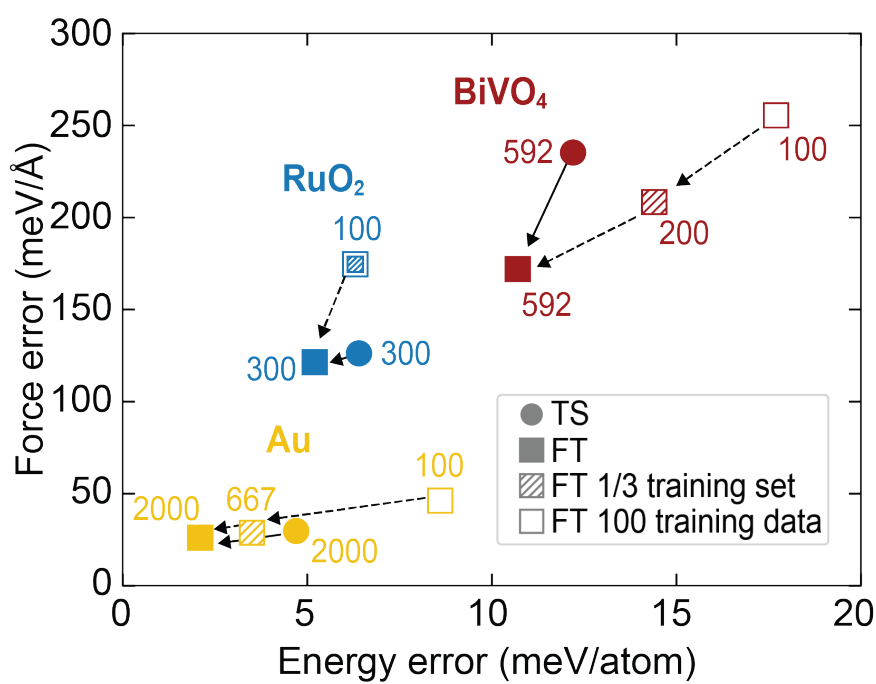

**Figure 5.** Prediction results for the three different surface systems studied in this work, distinghuished by different colors (Au in yellow, $RuO_2$ in blue, and $BiVO_4$ in red). Root mean squared prediction errors of energies and forces on the test set are shown as a function of training data size and method. Two approaches (*TS* and *FT*) are represented by circle and square shapes, respectively.

**Discussion**

In this study, we investigated the effectiveness of fine-tuning strategies applied to foundation models for efficiently developing MLPs targeted at unary, binary, and ternary material surface systems. Our results consistently demonstrated that *FT* significantly reduces prediction errors compared to models trained without prior knowledge, yielding predictions that are both physically and chemically more accurate. These improvements are clearly illustrated in Figure 5, which compares prediction errors between *TS* models (circles) and those obtained by *FT* model (squares) across three surface systems: Au, $RuO_2$, and $BiVO_4$. The *FT* models consistently outperformed the *TS* models and achieve comparable prediction accuracy even when trained with only one-third of the dataset or as few as 100 data points. Furthermore, the comparison across these systems indicates that prediction performance gradually declines as system complexity increases, likely due to the more intricate interactions among the elemental species.

Considering both numerical prediction accuracy and physical realism (as assessed through diatomic potential analysis), the *FT* model exhibits clear advantages. While additional research is necessary to identify optimal additional data quantities and characteristics, our results indicate that fine-tuning significantly enhances predictive accuracy significantly with fewer training samples. Consequently, *FT* models exhibit broader generalization capabilities, improving predictions even for OOD scenarios. Our findings confirm the suitability of fine-tuned models for gold surface studies; however, it remains uncertain whether this approach can be effectively generalized to other novel metal surfaces. Preliminary attempts on other unary metals revealed that fine-tuning can sometimes negatively affect predictive performance. Case 4 addresses this issue in detail, analyzing underlying causes and exploring strategies to mitigate these limitations.

Although foundation models trained on bulk data are known to underestimate surface energies (softening) [12], our results indicate that the consistent force-level mispredictions observed on surfaces arise primarily from the absence of surface-specific training coverage in the foundation model—an issue effectively mitigated by fine-tuning. Figure 3 for $BiVO_4$ also reveals a *ZS*-specific bias: a dense low-force cluster around ≈0.4 eV/Å indicating overestimation, and a prevalence of over-predictions relative to under-predictions. We attribute this to missing surface coverage in pre-training; FT mitigates it in practice.

Our analysis in Case 4 further highlights that *MHFT* substantially enhances predictive accuracy on OOD data, effectively overcoming a limitation frequently observed in conventional *FT*. Accordingly, *MHFT* is particularly advantageous in scenarios where researchers must preserve the predictive performance of the original bulk-trained model while simultaneously improving accuracy on a small, highly specialized dataset—for example, extending a bulk-trained UIP to complex multicomponent surfaces, reactive intermediates, or rare-event configurations where data are scarce. To fully exploit these advantages, however, careful control of training conditions becomes essential.

A crucial insight from our investigation is the importance of aligning computational conditions and energy reference points between pre-training and fine-tuning datasets, especially when employing *MHFT*. Discrepancies arising from differing simulation conditions can result in inconsistent predictions under *MHFT*. By contrast, conventional *FT* exhibits greater robustness, effectively tolerating moderate variations in simulation conditions without substantial degradation of performance.

To offer practical guidelines, consider a research scenario where only a limited amount of first-principles data (on the order of tens of structures, based on our study) is available for the target system. In such cases, a foundation model can be supplemented with additional data closely related, though not identical, to the target system. Researchers can then compare the prediction accuracy of the resulting fine-tuned (*FT* or *MHFT*) model to the zero-shot foundation model to identify the optimal starting point. This approach significantly reduces computational costs during subsequent predictions, explorations, and analyses.

Finally, to clarify the underlying mechanisms and support future application of *FT* and *MHFT* strategies, Figure 6 provides a schematic illustration of a simplified two-dimensional representation of the material feature space, including pre-training data and novel unary metal surfaces. The zero-shot scenario demonstrates reasonable predictive capabilities even without explicit surface-system training. However, conventional *FT*, while enhancing accuracy within specifically fine-tuned regions, often sacrifices predictive reliability for the original pre-training data. In contrast, *MHFT* maintains foundational predictive abilities from pre-training data while simultaneously improving accuracy both within and beyond the fine-tuning region. Thus, *MHFT* offers robust performance across both original and new datasets.

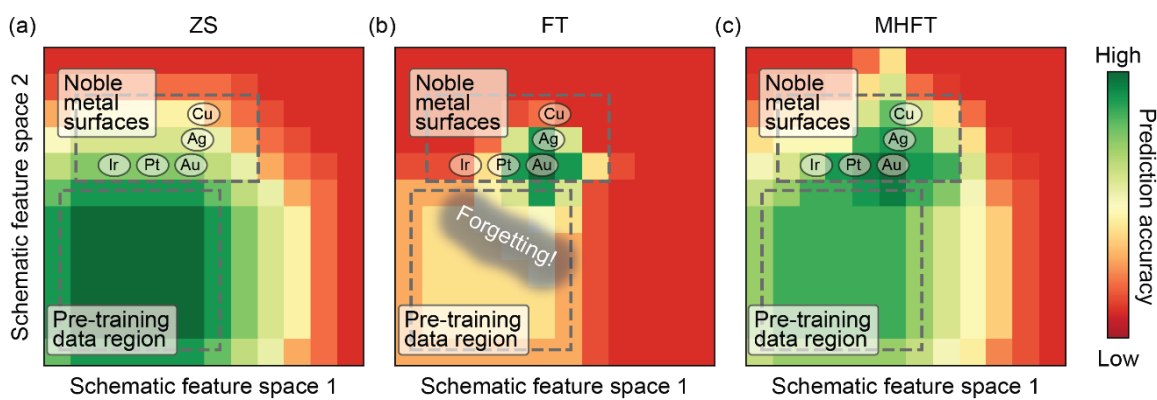

**Figure 6.** Schematic illustration of the prediction accuracy of three (a) zero-shot, (b) fine-tuning, and (c) multi-head fine-tuning models in a hypothetical materials feature space. Green regions represent high prediction accuracy for materials, while red indicates low. Various material systems are projected onto a simplified feature space: the bottom-left region represents bulk data used for pre-training the foundation model, and the upper region corresponds to noble metal surfaces. Elements adjacent to Au in the periodic table are also assumed to be positioned near Au in the feature space.

**Methodology**

All training and inference of the MACE potential were conducted using ACEsuit/MACE version 0.3.12. Unless stated otherwise, hyperparameters matched those of the foundation model. The key hyperparameters are summarized below, with their corresponding variable names provided in parentheses: evaluation metric (error_table) was per-atom RMSE, cutoff radius (r_max) was 6.0 Å, maximum spherical-harmonic expansion (max_ell) was 3, correction order (correction) was 3, number of message-passing interactions (num_interactions) was 2, activation function for the final readout (gate) was SiLU, batch size was 4, optimizer was Adam, and maximum learning rate (lr) was 0.02. The model was trained on both the total system energy and atomic forces, with the relative contributions to the loss function set to 10.0 for the energy term and 1.0 for the force term. Training minimized mean-squared error, while evaluation metrics were reported as per-atom RMSE.

For *FT* and *MHFT*, the learning rate was reduced to 0.001 while all other settings remained unchanged, to prevent rapid parameter drift and to ensure stable convergence of the pretrained weights on small domain-specific data. The *MHFT*-specific hyperparameter (weight_pt_head), which controls the relative contribution of the pre-training dataset loss to the overall loss, was set to 0.1. Larger values would favor predictive accuracy on the pre-training dataset; however, exhaustive hyperparameter optimization was not performed in this study.

During training, 20% of the available data (excluding the test set) was randomly selected as a validation set for early stopping. The remaining 80% was used exclusively for training, i.e., for parameter updates via back-propagation. For training the final model (i.e., the one using the largest dataset), the number of epochs required was as follows: for Gold, *FT/MHFT* (*TS*) required 20 (200) epochs; for $RuO_2$, both *FT* and *TS* required 250 epochs; and for $BiVO_4$, both *FT* and *TS* required 1,000 epochs. The checkpoint with the lowest validation loss was consistently selected as the final model, which was subsequently evaluated on the independent test set across the entire study.

For DFT calculations performed in this work, we employed VASP version 6.5.0 [44–46]. Isolated atoms and diatomic pairs were modeled in 20 × 21 × 22 Å orthorhombic cells and computed with spin polarization, sampling only the Γ point. The calculations used the generalized gradient approximation with the Perdew–Burke–Ernzerhof (GGA-PBE) functional [47], with an energy cutoff of 600 eV and Gaussian smearing with a width of 0.01 eV. To ensure consistency between our DFT calculations and the MACE predictions, we used the same VASP 5.4 PBE pseudopotentials as those used in the MPtrj dataset [42].

**Acknowledgements**

S.-H.Y. acknowledges support from the basic project from the Korea Research Institute of Chemical Technology in the Republic of Korea (KK2551-10). This research was supported by the Nano & Material Technology Development Program through the National Research Foundation of Korea(NRF) funded by Ministry of Science and ICT (RS-2024-00446683).

**Author contributions**

J.H. and S.-H.Y. designed the overall experiment. J.H. carried out the calculation under the supervision of S.-H.Y., analyzed the data with support from Y.L., T.L., and S.-H.Y. All authors drafted the manuscript and have given approval to the final version of the manuscript.

**Competing Interests**

The authors declare no competing interests.

**Data availability statement**

The data that supporting the findings of this study are available in the article and Supplementary Information or are available from the corresponding author upon reasonable request. Source data for all figures are provided as accompanying CSV files. Source data are provided with this paper.

**Code availability statement**

Custom scripts used in this study are available from the corresponding author upon reasonable request.

# References

[1] Behler, J. & Parrinello, M. Generalized neural-network representation of high-dimensional potential-energy surfaces. *Phys. Rev. Lett.* **98**, 146401 (2007).

[2] Bartók, A. P., Payne, M. C., Kondor, R. & Csányi, G. Gaussian approximation potentials: The accuracy of quantum mechanics, without the electrons. *Phys. Rev. Lett.* **104**, 136403 (2010).

[3] Behler, J. Perspective: Machine learning potentials for atomistic simulations. *J. Chem. Phys.* **145**, 170901 (2016).

[4] Deringer, V. L., Caro, M. A. & Csányi, G. Machine learning interatomic potentials as emerging tools for materials science. *Adv. Mater.* **31**, 1902765 (2019).

[5] Becker, C. A., Tavazza, F., Trautt, Z. T. & Buarque de Macedo, R. A. Considerations for choosing and using force fields and interatomic potentials in materials science and engineering. *Curr. Opin. Solid State Mater. Sci.* **17**, 277–283 (2013).

[6] Hale, L. M., Trautt, Z. T. & Becker, C. A. Evaluating variability with atomistic simulations: the effect of potential and calculation methodology on the modeling of lattice and elastic constants. *Model. Simul. Mater. Sci. Eng.* **26**, 055003 (2018).

[7] Center for Theoretical and Computational Materials Science (CTCMS), National Institute of Standards and Technology (NIST). Interatomic Potentials Repository. https://www.ctcms.nist.gov/potentials/, accessed 25 Sep 2025.

[8] Batzner, S. *et al.* E(3)-equivariant graph neural networks for data-efficient and accurate interatomic potentials. *Nat. Commun.* **13**, 2453 (2022).

[9] Batatia, I. *et al.* A foundation model for atomistic materials chemistry. *arXiv* preprint arXiv:2401.00096 (2024).

[10] Chen, C. & Ong, S. P. A universal graph deep learning interatomic potential for the periodic table. *Nat. Comput. Sci.* **2**, 718–728 (2022).

[11] Dunn, A., Wang, Q., Ganose, A., Dopp, D. & Jain, A. Benchmarking materials property prediction methods: the Matbench test set and Automatminer reference algorithm. *npj Comput. Mater.* **6**, 138 (2020).

[12] Focassio, B., Freitas, L. P. M. & Schleder, G. R. Performance assessment of universal machine learning interatomic potentials: challenges and directions for materials' surfaces. *ACS Appl. Mater. Interfaces* **17**, 13111–13121 (2025).

[13] Liu, X. *et al.* Fine-tuning universal machine-learned interatomic potentials: a tutorial on methods and applications. *arXiv* preprint arXiv:2506.21935 (2025).

[14] Tajbakhsh, N. *et al.* Convolutional neural networks for medical image analysis: full training or fine tuning? *IEEE Trans. Med. Imaging* **35**, 1299–1312 (2016).

[15] Lin, X. *et al.* Data-efficient fine-tuning for LLM-based recommendation. In *Proceedings of the 47th International ACM SIGIR Conference on Research and Development in Information Retrieval (SIGIR '24)*, 365–374 (2024).

[16] Radova, M., Stark, W. G., Allen, C. S., Maurer, R. J. & Bartók, A. P. Fine-tuning foundation models of materials interatomic potentials with frozen transfer learning. *npj Comput. Mater.* **11**, 237 (2025).

[17] Kim, J. *et al.* An efficient forgetting-aware fine-tuning framework for pretrained universal machine-learning interatomic potentials. *arXiv* preprint arXiv:2506.15223 (2025).

[18] French, R. M. Catastrophic forgetting in connectionist networks. *Trends Cogn. Sci.* **3**, 128–135 (1999).

[19] Church, K. W., Chen, Z. & Ma, Y. Emerging trends: a gentle introduction to fine-tuning. *Nat. Lang. Eng.* **27**, 763–778 (2021).

[20] Szép, M., Rueckert, D., von Eisenhart-Rothe, R. & Hinterwimmer, F. A practical guide to fine-tuning language models with limited data. *arXiv* preprint arXiv:2411.09539 (2024).

[21] Ruder, S. An overview of multi-task learning in deep neural networks. *arXiv* preprint arXiv:1706.05098 (2017).

[22] Stickland, A. C. & Murray, I. BERT and PALs: projected attention layers for efficient adaptation in multi-task learning. In *Proceedings of the 36th International Conference on Machine Learning (ICML 2019)*, PMLR **97**, 5986–5995 (2019).

[23] Park, Y., Kim, J., Hwang, S. & Han, S. Scalable parallel algorithm for graph neural network interatomic potentials in molecular dynamics simulations. *J. Chem. Theory Comput.* **20**, 4857–4868 (2024).

[24] Batatia, I., Kovács, D. P., Simm, G. N. C., Ortner, C. & Csányi, G. MACE: higher order equivariant message passing neural networks for fast and accurate force fields. *Adv. Neural Inf. Process. Syst.* **35**, 11423–11436 (2022).

[25] Neumann, M. *et al.* Orb: a fast, scalable neural network potential. *arXiv* preprint arXiv:2410.22570 (2024).

[26] Liao, Y.-L., Wood, B., Das, A. & Smidt, T. EquiformerV2: improved equivariant transformer for scaling to higher-degree representations. In *Proceedings of the International Conference on Learning Representations (ICLR 2024)* (2024).

[27] Haruta, M. Gold as a novel catalyst in the 21st century: preparation, working mechanism and applications. *Gold Bull.* **37**, 27–36 (2004).

[28] Barth, J. V., Brune, H., Ertl, G. & Behm, R. J. Scanning tunneling microscopy observations on the reconstructed Au(111) surface: atomic structure, long-range superstructure, rotational domains, and surface defects. *Phys. Rev. B* **42**, 9307–9318 (1990).

[29] Owen, C. J., Xie, Y., Johansson, A., Sun, L. & Kozinsky, B. Low-index mesoscopic surface reconstructions of Au surfaces using Bayesian force fields. *Nat. Commun.* **15**, 3790 (2024).

[30] Timmermann, J. et al. IrO$_2$ surface complexions identified through machine learning and surface investigations. *Phys. Rev. Lett.* **125**, 206101 (2020).

[31] Timmermann, J. et al. Data-efficient iterative training of Gaussian approximation potentials: application to surface structure determination of rutile IrO$_2$ and RuO$_2$. *J. Chem. Phys.* **155**, 244107 (2021).

[32] Lee, Y. et al. Staged training of machine-learning potentials from small to large surface unit cells: efficient global structure determination of the RuO$_2$ (100)-c(2 × 2) reconstruction and (410) vicinal. *J. Phys. Chem. C* **127**, 17599–17608 (2023).

[33] Lee, Y., Timmermann, J., Panosetti, C., Scheurer, C. & Reuter, K. Staged training of machine-learning potentials from small to large surface unit cells: efficient global structure determination of the RuO$_2$(100)-c(2 × 2) reconstruction and (410) vicinal. *J. Phys. Chem. C* **127**, 17599–17608 (2023).

[34] Owen, C. J. et al. Complexity of many-body interactions in transition metals via machine-learned force fields from the TM23 data set. *npj Comput. Mater.* **10**, 92 (2024).

[35] Liu, J., Byggmästar, J., Fan, Z., Qian, P. & Su, Y. Large-scale machine-learning molecular dynamics simulation of primary radiation damage in tungsten. *Phys. Rev. B* **108**, 054312 (2023).

[36] van der Oord, C. et al. Hyperactive learning for data-driven interatomic potentials. *npj Comput. Mater.* **9**, 168 (2023).

[37] Timmermann, J., Lee, Y., Staacke, C. G., Margraf, J. T., Scheurer, C. & Reuter, K. Data-efficient iterative training of Gaussian approximation potentials: application to surface structure determination of rutile IrO$_2$ and RuO$_2$. *J. Chem. Phys.* **155**, 244107 (2021).

[38] Lee, Y. & Lee, T. Machine-learning-accelerated surface exploration of reconstructed BiVO$_4$(010) and characterization of their aqueous interfaces. *J. Am. Chem. Soc.* **147**, 7799–7808 (2025).

[39] Kirkpatrick, J. *et al.* Overcoming catastrophic forgetting in neural networks. *Proc. Natl Acad. Sci. USA* **114**, 3521–3526 (2017).

[40] Li, Z. & Hoiem, D. Learning without forgetting. *IEEE Trans. Pattern Anal. Mach. Intell.* **40**, 2935–2947 (2018).

[41] Tran, R. *et al.* Surface energies of elemental crystals. *Sci. Data* **3**, 160080 (2016).

[42] Deng, B. *et al.* CHGNet as a pretrained universal neural network potential for charge-informed atomistic modelling. *Nat. Mach. Intell.* **5**, 1031–1041 (2023).

[43] Lejaeghere, K. *et al.* Reproducibility in density functional theory calculations of solids. *Science* **351**, 6280, aad3000 (2016).

[44] Kresse, G. & Furthmüller, J. Efficient iterative schemes for *ab initio* total-energy calculations using a plane-wave basis set. *Phys. Rev. B* **54**, 11169–11186 (1996).

[45] Kresse, G. & Furthmüller, J. Efficiency of *ab initio* total energy calculations for metals and semiconductors using a plane-wave basis set. *Comput. Mater. Sci.* **6**, 15–50 (1996).

[46] Kresse, G. & Joubert, D. From ultrasoft pseudopotentials to the projector augmented-wave method. *Phys. Rev. B* **59**, 1758–1775 (1999).

[47] Perdew, J. P., Burke, K. & Ernzerhof, M. Generalized gradient approximation made simple. *Phys. Rev. Lett.* **77**, 3865–3868 (1996).

**Supplementary Information for "Fine-Tuning Bulk-oriented Universal Interatomic Potentials for Surfaces: Accuracy, Efficiency, and Forgetting Control"**

"

**Jaekyun Hwang1, Taehun Lee2, Yonghyuk Lee3,\*, Su-Hyun Yoo1,\***

1. Digital Chemical Research Center, Korea Research Institute of Chemical Technology, Daejeon 34114, Republic of Korea
2. Division of Advanced Materials Engineering, Jeonbuk National University, Jeonju 54896, Republic of Korea
3. Department of Chemical and Biochemical Engineering, Dongguk University, Seoul, 04620, Republic of Korea

\* yhyuk@dongguk.edu, syoo@krict.re.kr

**Supplementary Notes 1. Benchmark Evaluation and Selection of Foundation Model**

To systematically investigate fine-tuning strategies for surface modeling, it is essential to first select a suitable universal interatomic potential (UIP) model as our foundation. Recent progress in assembling extensive open databases, such as MPtrj[1], Alexandria[2], and OMat24[3], has facilitated the development of comprehensive bulk-based UIPs trained on vast first-principles calculation datasets. Despite their proven accuracy in bulk predictions, these UIPs still show suboptimal performance when predicting surface energies and other out-of-domain scenarios [4,5]. Given these limitations, we first conducted a benchmark evaluation to determine the predictive robustness, generalization capability, and physical reliability of several pre-trained UIP models, thus guiding the selection of an appropriate foundational model.

Diatomic potentials describe how the potential energy between two isolated atoms changes as their separation distance decreases from infinity to shorter distances in vacuum. Although such scenarios typically fall outside the scope of bulk-oriented training datasets, they provide valuable benchmarks for evaluating the predictive robustness and extrapolation capabilities of UIPs. Despite their limited direct relevance to bulk or surface energy predictions, the simplicity and intuitive nature of diatomic interactions make them an effective benchmark for evaluating model generalizability to previously unseen atomic configurations.

A robust and well-trained UIP should accurately capture both short-range repulsive interactions and long-range attractive interactions, resulting in a smooth potential energy curve analogous to a classical Lennard-Jones potential with a clearly defined energy minimum [6]. To evaluate this capability, we compared the predicted diatomic potentials of five pre-trained UIP models against reference DFT calculations performed using the Perdew–Burke–Ernzerhof (PBE) functional [7]. Supplementary Fig. 1 explicitly shows this comparison for the hydrogen dimer ($H_2$).

All UIP models evaluated in this benchmark were tested in a zero-shot setting, meaning they received no additional training specific to diatomic interactions. Supplementary Table 1 summarizes detailed characteristics of these models, including training datasets, the number of training structures, and model complexities.

Most UIPs predicted equilibrium bond lengths consistent with the DFT reference values. Although most UIPs qualitatively reproduced the overall shape of the potential energy curves, many models exhibited irregularities or "bumps" (1.7 Å, 1.9 Å, and 2.2 Å from ORBv2, Eqv2, and M3GNet, respectively.) suggesting a lack of smoothness. Specifically, ORBv2 and eqV2, attention-based equivariant graph neural networks trained on large-scale datasets, presented notably rough potential energy surfaces. Despite achieving strong performance on standard benchmarks such as Matbench-discovery [8], these irregularities may undermine confidence in their application, particularly in molecular dynamics simulations where smooth and physically consistent potentials are essential. Additionally, Supplementary Fig. 2 shows nine homonuclear diatomic potentials, including those for the elements used in the subsequent case studies. Although a quantitative comparison is difficult, the MACE model demonstrates consistently superior performance without unphysical behavior such as discontinuities or multiple energy minima.

Given that these UIP models were trained exclusively on bulk-oriented datasets, their performance in predicting isolated diatomic pair potentials is notably encouraging. Specifically, MACE and SevenNet demonstrated commendable accuracy and consistency compared to the DFT reference data, highlighting their versatility for applications beyond bulk simulations.

Based on the benchmark results, the MACE model slightly underestimated the binding energy but exhibited physically consistent behavior and high accuracy near the equilibrium region. The curvature and smoothness of its potential energy curve were especially noteworthy. Although all UIP foundation models were trained on bulk database, we anticipated that MACE would be suitable to test fine-tuning beyond bulk domains considering its predictive accuracy, the smoothness of its potential energy curves, and the availability of ongoing software support. Despite not achieving the top performance on bulk benchmarks, we expect MACE as the optimal candidate among the UIP foundation models for our study.

For all subsequent zero-shot and fine-tuning experiments described in this study, we employed the same pre-trained MACE checkpoint evaluated in the diatomic benchmark. This methodological choice ensures consistency and enables clear comparative analysis.

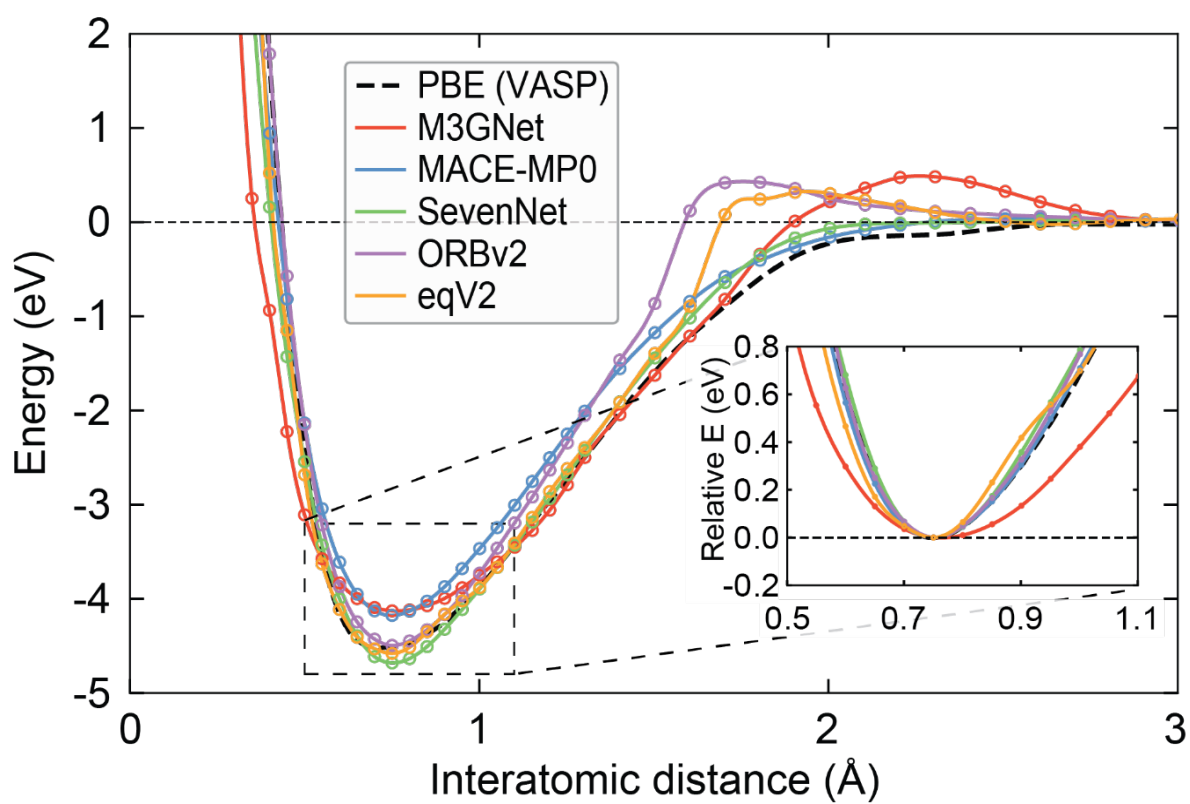

**Supplementary Figure 1.** Comparison of diatomic pair potential energy curves for the hydrogen dimer ($H_2$) predicted by various pre-trained UIPs and DFT-PBE reference calculations. All energy curves are corrected to have zero energy at infinite atomic separation. The inset povides a magnified view near the equilibrium bond length (~0.75 Å), where the relative energies are referenced to the respective potential minima.

| Name | Training dataset | Number of training structures | Number of parameters | Model keywords |
|---|---|---|---|---|
| M3GNet [9] | MPF.2021.2.8 | 187,687 | 227,549 | M3GNet-MP-2021.2.8-PES |
| SevenNet [10] | MPTrj | 1,580,395 | 1,171,144 | 7net-l3i5 |
| MACE [4,11] | MPTrj | 1,580,395 | 3,847,696 | 2023-12-10-mace-128-L0_energy_epoch-249 |
| ORBv2 [12] | MPTrj & Alex | 32,078,023 | 25,161,727 | orb-v2-20241011 |
| eqV2 [13] | MPTrj & OMat24 | 102,404,980 | 86,589,068 | eqV2_86M_omat_mp_salex |

**Supplementary Table 1.** Summary of the UIP models investigated in this benchmarking study.

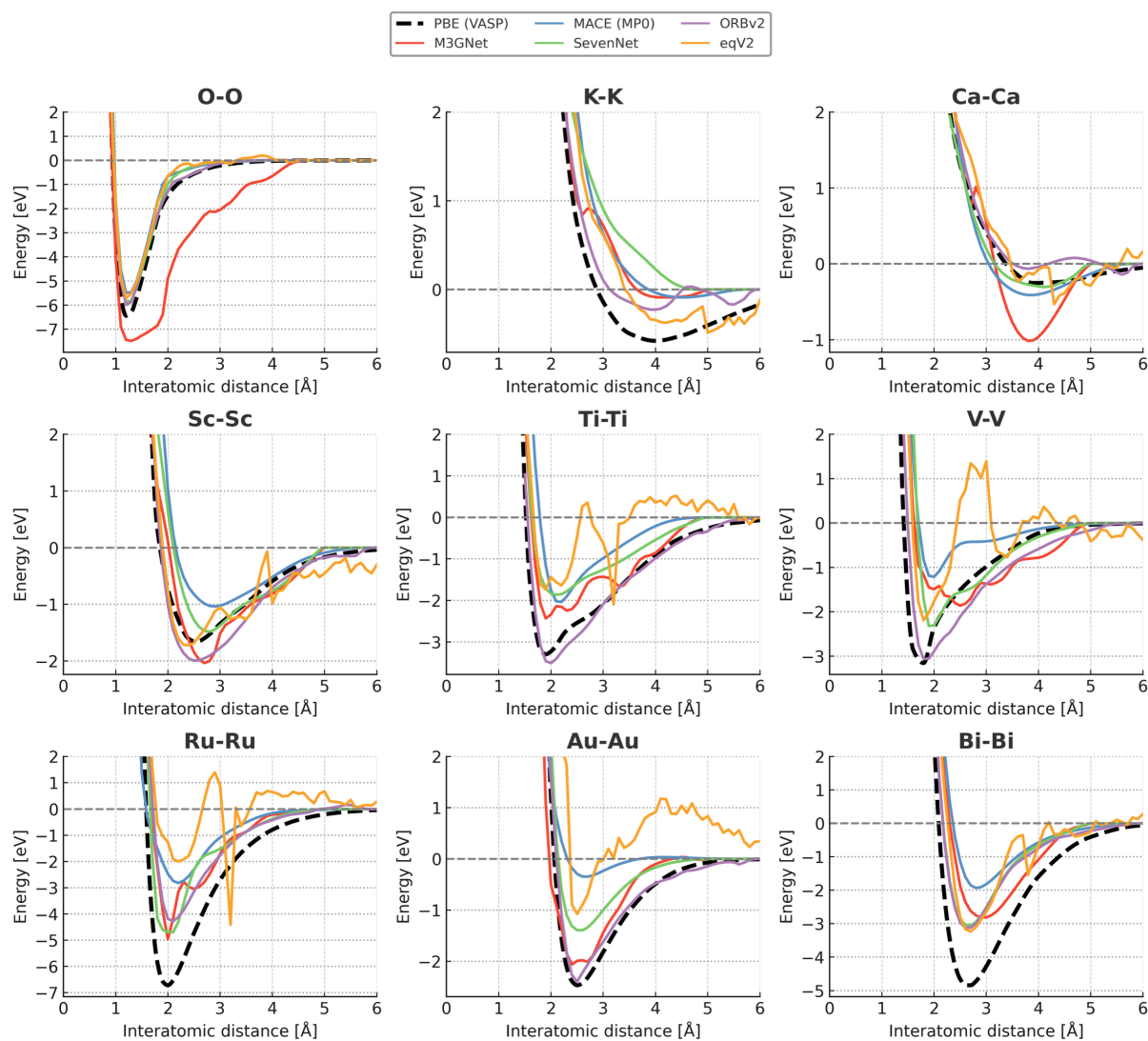

**Supplementary Figure 2.** Homonuclear diatomic pair potential energy curves for nine elements predicted by various pre-trained UIPs and DFT-PBE reference calculations identical to Supplementary Figure 1. All energy curves are corrected to have zero energy at infinite atomic separation.

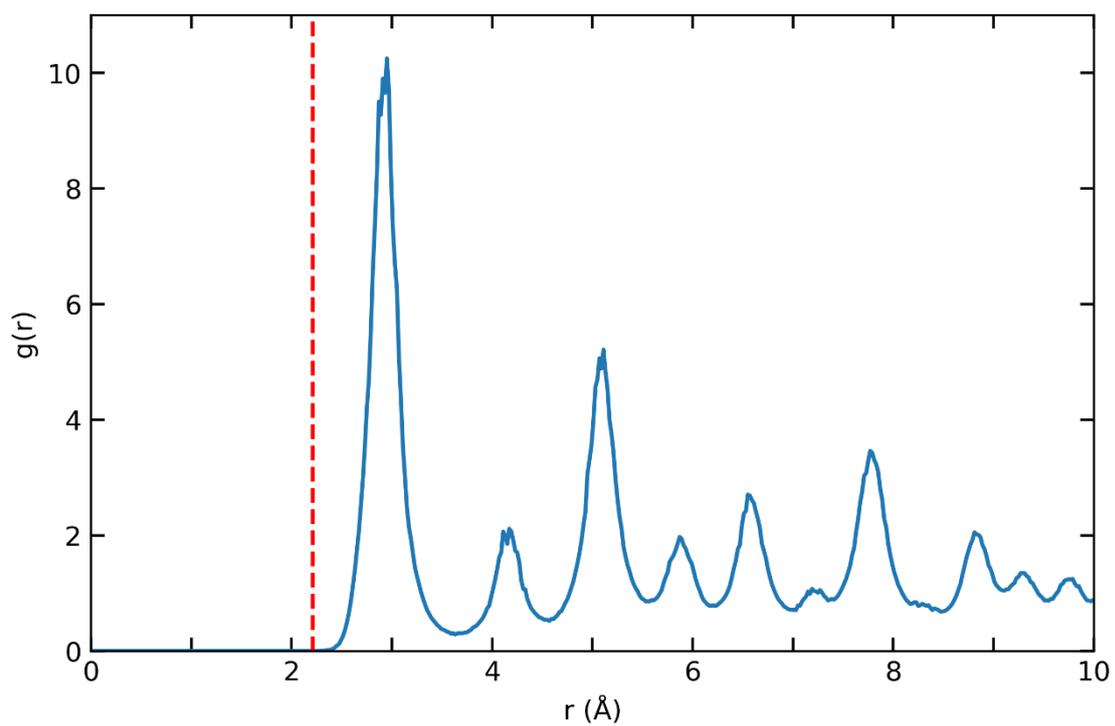

**Supplementary Figure 3.** Au-Au radial distribution of Au training dataset. No Au-Au bond exists in datasets shorther than 2.21Å

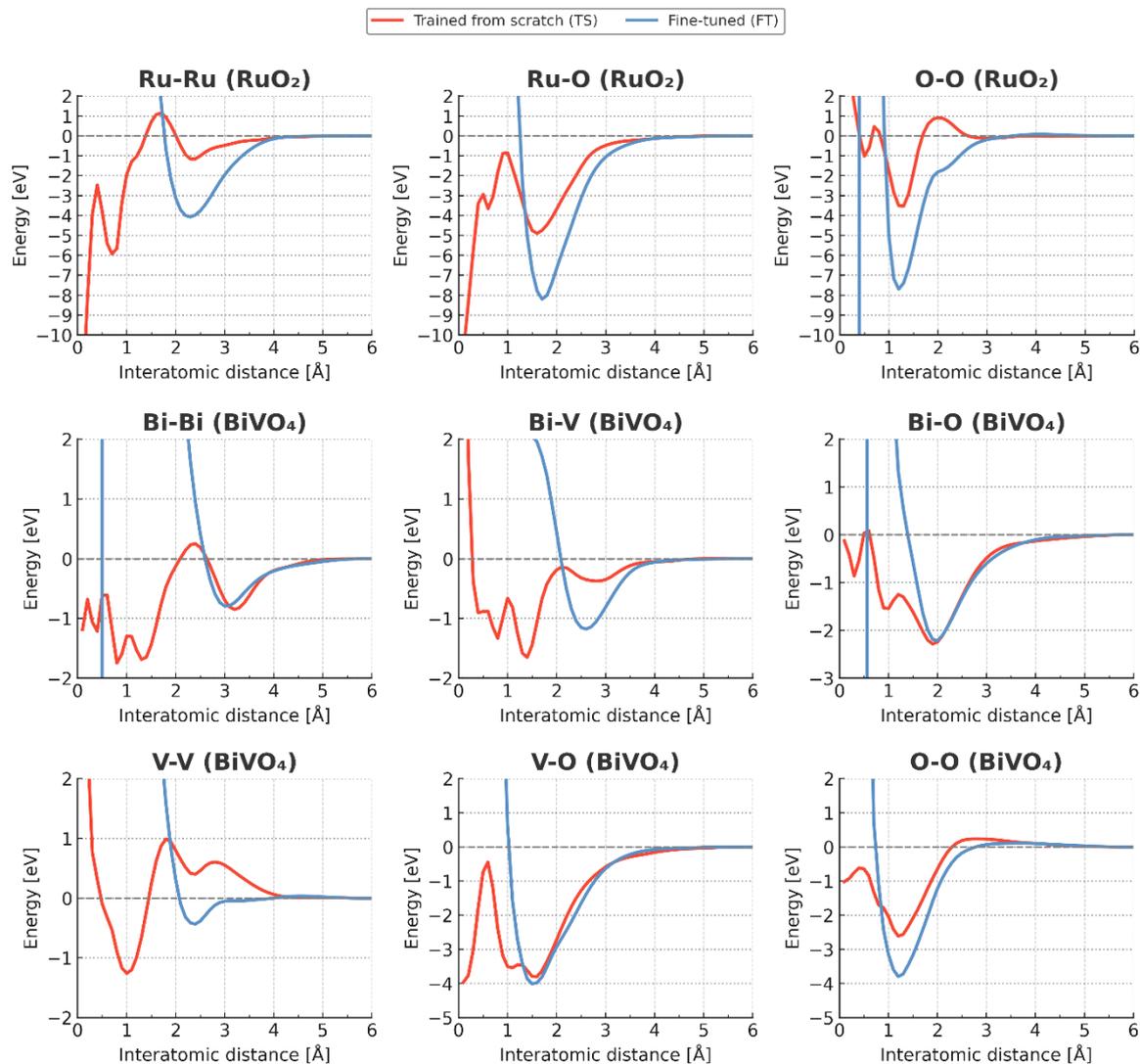

**Supplementary Figure 4.** Predicted potential energy curves for six possible diatomic pairs in the $BiVO_4$ ternary system. The graph show the variation in energy as a function of interatomic distance, as predicted by the TS and FT models.

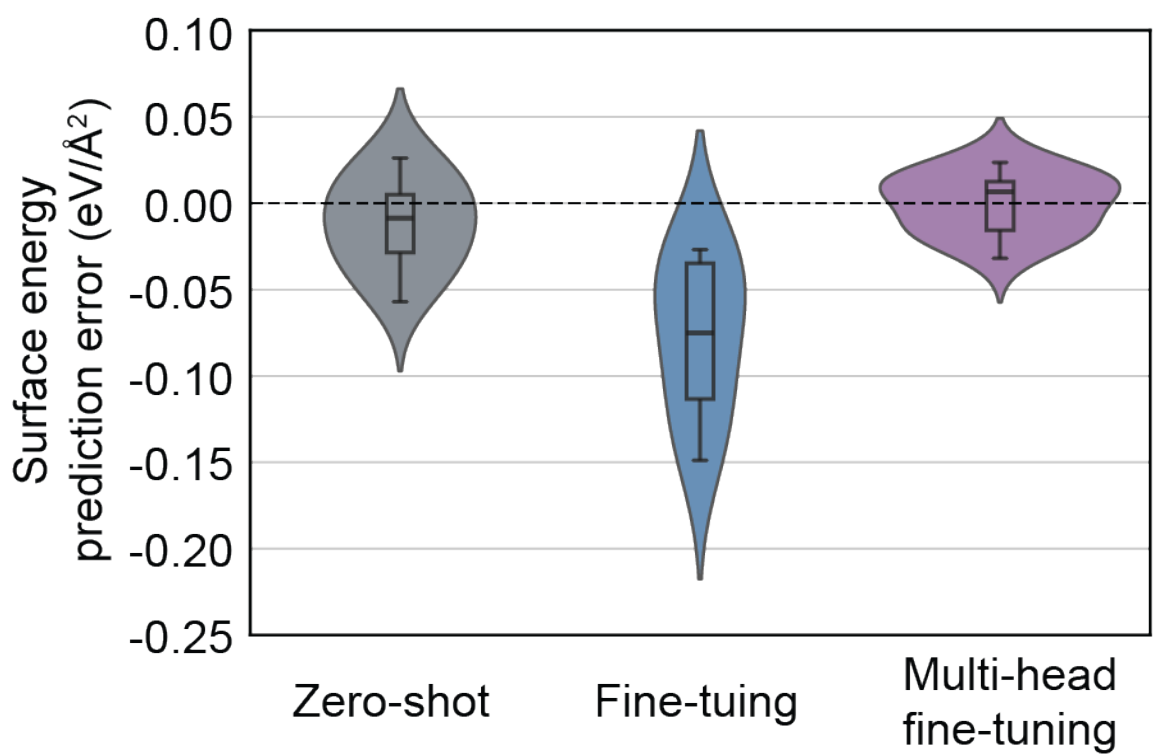

**Supplementary Figure 5.** Violin plots of prediction errors in surface energy for surfaces of four noble metal elements (Ag, Cu, Pt, Ir) from three different models. Training was performed using 2,000 Au surfaces for *FT* and *MHFT* models, while all predictions were performed on 54 surfaces of Ag, Cu, Pt, Ir, excluding Au. The areas of the violins are normalized to be equal.

# References

[1] French, R. M. Catastrophic forgetting in connectionist networks. *Trends Cogn. Sci.* **3**, 128–135 (1999).

[2] Church, K. W., Chen, Z. & Ma, Y. Emerging trends: a gentle introduction to fine-tuning. *Nat. Lang. Eng.* **27**, 763–778 (2021).

[3] Szép, M., Rueckert, D., von Eisenhart-Rothe, R. & Hinterwimmer, F. A practical guide to fine-tuning language models with limited data. *arXiv* preprint arXiv:2411.09539 (2024).

[4] Batatia, I. *et al.* A foundation model for atomistic materials chemistry. *arXiv* preprint arXiv:2401.00096 (2024).

[5] Focassio, B., Freitas, L. P. M. & Schleder, G. R. Performance assessment of universal machine learning interatomic potentials: challenges and directions for materials' surfaces. *ACS Appl. Mater. Interfaces* **17**, 13111–13121 (2025).

[6] Ruder, S. An overview of multi-task learning in deep neural networks. *arXiv* preprint arXiv:1706.05098 (2017).

[7] Perdew, J. P., Burke, K. & Ernzerhof, M. Generalized gradient approximation made simple. *Phys. Rev. Lett.* **77**, 3865–3868 (1996).

[8] Dunn, A., Wang, Q., Ganose, A., Dopp, D. & Jain, A. Benchmarking materials property prediction methods: the Matbench test set and Automatminer reference algorithm. *npj Comput. Mater.* **6**, 138 (2020).

[9] Chen, C. & Ong, S. P. A universal graph deep learning interatomic potential for the periodic table. *Nat. Comput. Sci.* **2**, 718–728 (2022).

[10] Park, Y., Kim, J., Hwang, S. & Han, S. Scalable parallel algorithm for graph neural network interatomic potentials in molecular dynamics simulations. *J. Chem. Theory Comput.* **20**, 4857–4868 (2024).

[11] Batatia, I., Kovács, D. P., Simm, G. N. C., Ortner, C. & Csányi, G. MACE: higher order equivariant message passing neural networks for fast and accurate force fields. *Adv. Neural Inf. Process. Syst.* **35**, 11423–11436 (2022).

[12] Neumann, M. *et al.* Orb: a fast, scalable neural network potential. *arXiv* preprint arXiv:2410.22570 (2024).

[13] Liao, Y.-L., Wood, B., Das, A. & Smidt, T. EquiformerV2: improved equivariant transformer for scaling to higher-degree representations. In *Proceedings of the International Conference on Learning Representations (ICLR 2024)* (2024).